\documentclass[11pt]{article}

\title{Subspace Clustering via Thresholding and Spectral Clustering}

\author{Reinhard Heckel and Helmut B\"olcskei
 \\[0.5em]
  \multicolumn{1}{p{.7\textwidth}}{\centering Dept. of IT \& EE, ETH Zurich, Switzerland}}

\date{}
\usepackage{setspace}

\renewcommand{\vspace}[1]{}

\usepackage{amssymb}

\usepackage[figuresright]{rotating}

\usepackage{amsmath,amssymb}
\usepackage{graphicx}
\usepackage{dcolumn}
\usepackage{bm}
\usepackage{amscd,amsthm}
\usepackage{ifthen}

\usepackage{mathtools} 

\usepackage[applemac]{inputenc}

\usepackage{tikz}
\usepackage{pgfplots}
\usepackage{subfigure}
\usetikzlibrary{pgfplots.groupplots}

\usepackage{bm}

\newcommand\norm[2][\Tnorm]{\ensuremath{{\left\Vert #2 \right\Vert}_{#1}}}

\newcommand\Tinnerprod{}
\newcommand{\innerprod}[3][\Tinnerprod]{\ifthenelse{\equal{#1}{}}{\ensuremath{\left<#2,#3\right>}}{\ensuremath{\left<#2,#3\right>_{#1}}}}

\newcommand\vect[1]{\mathbf #1}

\newcommand{\US}[1]{S^{#1-1}} 

\newcommand\Tex{}
\newcommand\EX[2][\Tex]{
\ifthenelse{\equal{#1}{}}{{\mathbb E}\left[#2\right]}{\ensuremath{{\mathbb E}_{#1}\left[ #2\right]}}}

\newcommand\Var[2][\Tex]{
\ifthenelse{\equal{#1}{}}{{\mathrm{Var} }[#2]}{\ensuremath{\mathrm{Var}_{#1}\left[ #2\right]}}}

\newcommand\ignore[1]{}

\newcommand\defeq{\coloneqq }

\newcommand{\reals}{\mathbb R} 
 
\newtheorem{definition}{Definition}

\newtheorem{theorem}{Theorem}

\newcommand{\herm}[1]{{#1}^T} 

\usepackage{hyperref}
\hypersetup{
    colorlinks=false,       
    linkcolor=red,          
    citecolor=green,        
    filecolor=magenta,      
    urlcolor=cyan           
}

\newcommand{\cS}{S} 
\newcommand{\X}{\mathcal X} 

\usepackage{bm}

\newcommand{\va}{\vect{a}}  
\newcommand{\vb}{\vect{b}}

\newcommand{\vv}{\vect{v}}  

\newcommand{\vx}{\vect{x}}  
  
\newcommand{\vz}{\vect{z}}

\newcommand{\mA}{\vect{A}}

\newcommand{\mI}{\vect{I}}

\newcommand{\mU}{\vect{U}}

\renewcommand{\S}{\mathcal T}

\newcommand{\D}{\mathcal D}

\renewcommand{\O}{\mathcal O}

\newcommand{\q}{q}

\begin{document}
\maketitle

\begin{abstract}
We consider the problem of clustering a set of high-dimensional data points into sets of low-dimensional linear subspaces. 
The number of subspaces, their dimensions, and their orientations are unknown. 
We propose a simple and low-complexity clustering algorithm based on thresholding the correlations between the data points followed by spectral clustering. A probabilistic performance analysis shows that this algorithm succeeds even when the subspaces intersect, and when the dimensions of the subspaces scale (up to a log-factor) linearly in the ambient dimension. Moreover, we prove that the algorithm also succeeds for data points that are subject to erasures with the number of erasures scaling (up to a log-factor) linearly in the ambient dimension. Finally, we propose a simple scheme that provably detects outliers. 
\end{abstract}



\vspace{-0.2cm}
\section{Introduction}
\label{sec:intro}
\vspace{-0.2cm}
Suppose we are given a set $\X$ of $N$ data points in $\reals^m$, and assume that $\X = \X_1 \cup ...  \cup  \X_L\cup \O$, where the points in $\X_l$ lie in a (low-dimensional) linear subspace $\cS_l$  of $\reals^m$, and $\O$ denotes a set of outliers. The association of the data points with the sets $\X_l$ and $\O$, the number of subspaces $L$, their dimensions $d_l$, and their orientations are all unknown. We consider the problem of clustering the data points, i.e., of finding the assignments of the points in $\X$ to the sets $\X_l$ and $\O$. 
Note that once these associations have been identified, it is straightforward to extract the subspaces $\cS_l$ through principal component analysis (PCA). 
The problem we consider is known as subspace clustering and has applications in, e.g., unsupervised learning, image processing, disease detection, and, in particular, computer vision, see, e.g., \cite{vidal_subspace_2011} and references therein. 
Numerous approaches to subspace clustering are known. 
We refer to \cite{vidal_subspace_2011} for an excellent overview. 

Spectral clustering (SC) methods (see \cite{luxburg_tutorial_2007} for an introduction) have found particularly widespread use. 
At the heart of SC lies the construction of an adjacency matrix $\mA \in \reals^{N\times N}$, with the $(i,j)$th entry of $\mA$ measuring the similarity between the data points $\vx_i, \vx_j \in \X$. A typical similarity measure is, e.g., $e^{-\mathrm{dist}(\vx_i,\vx_j)}$, where $\mathrm{dist}(\cdot,\cdot)$ is some distance measure \cite{vidal_subspace_2011}. 
Taking $G$ to be the graph with adjacency matrix $\mA$, the association of the points in $\X$ to the sets $\X_l$ (outliers are typically removed in a preprocessing step) is obtained by finding the connected components in $G$, accomplished via singular value decomposition of the Laplacian of $G$ followed by k-means clustering \cite{luxburg_tutorial_2007}. 
Whether a SC algorithm, or for that matter, any clustering algorithm, succeeds 
depends on the number of subspaces $L$, their dimensions and relative orientations, and the number of points in each subspace. 
Analytic results on the performance of SC methods are scarce. A notable exception is the \emph{sparse subspace clustering} (SSC) algorithm, recently introduced by Elhamifar and Vidal \cite{elhamifar_sparse_2009,elhamifar_sparse_2012}. At the heart of this algorithm lies a clever construction of $\mA$ that uses ideas from sparse signal recovery. Soltanolkotabi and Cand\`es \cite{soltanolkotabi_geometric_2011} presented an elegant (geometric function) analysis of SSC and proved that SSC succeeds under very general conditions. 
Most importantly, it is shown in \cite{soltanolkotabi_geometric_2011}, using a probabilistic analysis, that SSC succeeds even when the subspaces $\cS_l$ intersect, which means the $\cS_l$ do not need to be independent or disjoint\footnote{The linear subspaces $\cS_l$ are called disjoint if $\cS_l \cap \cS_k = \{\vect{0}\}$ for all $l\neq k$, and independent if $\dim ( \oplus_l \cS_l ) = \sum_l d_l$, where $\oplus$ stands for direct sum. 
An independent set of subspaces is disjoint, but the converse is not necessarily true. Two subspaces are said to intersect if $\cS_l \cap \cS_k \neq \{\vect{0}\}$. 
}. Moreover, Soltanolkotabi and Cand\`es \cite{soltanolkotabi_geometric_2011} provide a clever extension of SSC that provably detects outliers. 
To construct the adjacency matrix $\mA$ SSC requires the solution of $N$ $\ell_1$-minimization problems, each in $N$ unknowns; this can pose significant computational challenges for large data sets.

{\bf Contributions:}
We introduce an algorithm, termed thresholding based subspace clustering (TSC), which applies spectral clustering to an adjacency matrix $\mA$ obtained by thresholding correlations between the data points in $\X$. 
 TSC is shown to succeed even when the subspaces intersect, and when their dimensions scale (up to a log-factor) linearly in the ambient dimension. 
While SSC shares these desirable properties, TSC is computationally much less demanding, as the construction of the adjacency matrix $\mA$ in the TSC algorithm requires the computation of $N^2$ inner products followed by thresholding only.  
Moreover, the performance analysis of TSC, thanks to the algorithm's simplicity, does not need sophisticated mathematical tools; it is based on fairly standard concentration results for order statistics only. 

In practical applications the data points to be clustered are often subject to erasures, caused by, e.g., scratches on images. The literature is essentially void of corresponding analytic performance results. We prove that TSC succeeds even when the data points in $\X$ are subject to massive erasures. Specifically, the number of erasures is allowed to scale (up to a log-factor) linearly in the ambient dimension. 
We finally propose a simple scheme that provably detects outliers, and we corroborate our findings by numerical results. 
Proofs of the theorems in this paper, results on clustering noisy data points, and numerical results for real data sets are provided in \cite{heckel_subspace_journal_2013}. 

We finally note that Lauer and Schnorr \cite{lauer_spectral_2009} also apply SC to an adjacency matrix constructed from correlations between data points, albeit, without thresholding. Moreover, no analytic performance results are available for the algorithm in \cite{lauer_spectral_2009}. 

{\bf Notation:} We use lowercase boldface letters to denote (column) vectors, e.g., $\vx$, and uppercase boldface letters to designate matrices, e.g., $\mA$. For the vector $\vx$, $[\vx]_q$ and $x_q$ denote the $q$th entry and for the matrix $\mA$, $\mA_{ij}$ stands for the entry in the $i$th row and $j$th column. 
The spectral norm of $\mA$ is $\norm[2\to 2]{\mA} \defeq\;$ $\max_{\norm[2]{\vv} = 1  } \norm[2]{\mA \vv}$, its Frobenius norm is $\norm[F]{\mA} \defeq \sqrt{\sum_{i,j} (\mA_{ij})^2 }$, and $\mI$ denotes the identity matrix. 
The superscript $\herm{}$ stands for transposition and $\log(\cdot)$ for the natural logarithm. 
 The cardinality of the set $\S$ is $|\S|$. 
We write $\mathcal N( \boldsymbol{\mu},\boldsymbol{\Sigma})$ for a Gaussian random vector with mean $\boldsymbol{\mu}$ and covariance matrix $\boldsymbol{\Sigma}$. 
The unit sphere in $\reals^m$ is $\US{m} \,\defeq\, \{ \vx \in \reals^m \colon \norm[2]{\vx} = 1 \}$. 

\vspace{-0.45cm}
\section{The TSC algorithm}
\vspace{-0.35cm}

The formulation introduced below assumes that outliers have already been removed from $\X$, e.g., through the outlier detection scheme in Sec.~\ref{sec:detoutl}. 
Given a set of data points\footnote{We assume the data points to be either normalized or to be of comparable norm. This assumption is not restrictive as the data points can be normalized prior to clustering. } $\X$ 
 and the parameter $q$ (the choice of $q$ is discussed below), the TSC algorithm consists of the following steps:

{\bf Step 1:} For every $\vx_j \in \X$, identify the set $\S_j \subset \{1,...,N\} \setminus j$ of cardinality $q$ such that 
\vspace{-0.1cm}
\begin{equation}
\left| \innerprod{\vx_j}{ \vx_i} \right| \geq \left| \innerprod{\vx_j}{ \vx_p} \right| \text{ for all } i \in \S_j \text{ and all } p \notin \S_j
\label{eq:thresholding}
\vspace{-0.1cm}
\end{equation}
and let $\vz_j \in \reals^N$ be the vector with $i$th entry $\left| \innerprod{\vx_j}{ \vx_i} \right|$ if $i\in \S_j$, and $0$ if $i\notin \S_j$. Construct the adjacency matrix according to $\mA_{ij} = |  [\vz_j]_i | + | [\vz_i]_j |$. 

{\bf Step 2:} Estimate the number of subspaces using the eigengap heuristic \cite{luxburg_tutorial_2007}  according to
$
\hat L =  \arg  \max_ {i=1,..., N-1}  \allowbreak(\lambda_{i+1} - \lambda_{i}),
$
where $\lambda_1\leq \lambda_2 \leq ... \leq \lambda_N$ are the eigenvalues of the normalized Laplacian of the graph with adjacency matrix $\mA$. 

{\bf Step 3:} Apply normalized SC \cite{luxburg_tutorial_2007} to $(\mA, \hat L)$.


\vspace{0.1cm}

TSC is said to succeed if the TSC subspace detection property according to the following definition holds.  
\vspace{-0.1cm}
\begin{definition}
The TSC subspace detection property holds for $\X = \X_1 \cup \,...  \,\cup \, \X_L$ and adjacency matrix $\mA$ if

\hspace{-0.1cm}{\it i.} $\mA_{ij} \neq0$ only if $\vx_i$ and $\vx_j$ belong to the same set $\X_l$

\noindent and if

\hspace{-0.1cm}{\it ii.} for all $i =1,...,N$, $\mA_{ij} \neq 0$ for at least $\q$ pairs $\vx_i$ and $\vx_j$ that belong to the same set $\X_l$.

\label{def:lsdp}
\end{definition}

\vspace{-0.3cm}
The idea behind Def.~\ref{def:lsdp}, inspired by the $\ell_1$ subspace detection property introduced in \cite{soltanolkotabi_geometric_2011}, is the following.
If the TSC subspace detection property holds, then each node in the Graph $G$ with adjacency matrix $\mA$ is connected to at least $\q$ other nodes, all of which correspond to points in the same subspace. In the SC step, the assignments of the points to clusters are then determined through identification of the connected components of $G$. 
We will see in the numerical results section, that even if the TSC subspace detection property does not hold strictly, but the $\mA_{ij}$ for pairs $\vx_i,\vx_j$ belonging to different subspaces are sufficiently small, SC can still yield the correct result.

{\bf Assumptions for performance analysis:}
For expositional convenience we take all subspaces to have equal dimension $d$, and let the number of points in each of the subspaces be $n$, (i.e., $|\X_l|=n, l=1,...,L$). 

{\bf Choice of $q$:}
Choosing $q$ too small/large will lead to over/under-estimation of the number of subspaces $L$. 
A sensible choice is to take $q$ to be a fraction of $n$. 
This motivates setting $q= n/ \rho$, where $\rho \geq 1$. The results we obtain will ensure that, under certain conditions, the TSC subspace detection property holds, provided that $\rho$ is not too small, while the specific choice of $\rho$ will not matter. Moreover, numerical results in \cite{heckel_subspace_journal_2013} indicate that TSC is not sensitive to the specific choice of $\q$.

\section{Deterministic subspaces}
\vspace{-0.35cm}

In order to understand the impact of the relative orientations of the subspaces on the performance of TSC, we take the subspaces to be deterministic and the points in the subspaces to be random. 
W.l.o.g.~we represent the points in $\cS_l$ as 
$
\vx^{(l)}_j = \mU^{(l)} \va^{(l)}_j
$, $j=1,...,n$, where $\va^{(l)}_j  \in \reals^{d}$ and $\mU^{(l)}$ is a basis for the $d$-dimensional subspace $\cS_l$. 
We present two results that depend on different notions of affinity between subspaces, namely 
\newcommand{\affp}{\mathrm{affp}}
\[
\affp(\cS_k,\cS_l) \defeq \norm[2\to 2]{ \herm{\mU^{(k)}} \mU^{(l)}  }
\]
and \cite[Def.~2.6]{soltanolkotabi_geometric_2011} 
\newcommand{\aff}{\mathrm{aff}}
\[
\aff(\cS_k,\cS_l) \defeq \norm[F]{ \herm{\mU^{(k)}} \mU^{(l)}  } /\sqrt{d},
\]
both of which can be interpreted as measures of the relative orientations of the subspaces. 
Throughout this section, we assume that the $\mU^{(l)}$ are orthonormal bases, and hence $0 \leq \aff(\cS_k,\cS_l)  \leq \affp(\cS_k,\cS_l) \leq 1$. 
The relation between the two affinity notions is brought out by noting that $\affp(\cS_k,\cS_l) = \cos(\theta_1)$ while $\aff(\cS_k,\cS_l) = \sqrt{ \cos^2( \theta_1) + ...+ \cos^2(\theta_d)}/\sqrt{d}$, where $\theta_1 \leq ... \leq \theta_d$ are the principal angles between $\cS_k$ and $\cS_l$. 
 
\vspace{-0.05cm}
\begin{theorem}
Suppose $n = \rho \q$ data points are chosen in each of the $L$ subspaces at random according to $\vx_j^{(l)} = \mU^{(l)} \va^{(l)}_j, j =1,..., n$, where the $\va^{(l)}_j$ are i.i.d.~$\mathcal N(0, \allowbreak(1/d)  \mI)$ and $\rho \geq 10/3$. If 
\vspace{-0.2cm}
\begin{align}
\max_{k \neq l} \affp(\cS_k,\cS_l) \leq c_1 \frac{\sqrt{ \log \rho }}{ \sqrt{\log L + \log n   }  },
\label{eq:condTSCaffpspec}
\end{align}
then the TSC subspace detection property holds with probability at least
$
1 - L n   e^{-c_2 n}  - \frac{L}{(L-1)^3 n^2 },
$
\label{thm:TSCprob}
where  $c_1$ and $c_2$ are absolute constants satisfying $0<c_1,c_2 <1$. 
\end{theorem}
\vspace{-0.1cm}

Thm.~\ref{thm:TSCprob} states that TSC succeeds with high probability 
if $\max_{k\neq l} \affp(\cS_k,\cS_l)$ is sufficiently small. Intuitively, we expect that 
clustering becomes easier when the number of data points in each subspace increases. Thm.~\ref{thm:TSCprob} confirms this intuition as, for fixed $d$, $\q$, and $L$, the right hand side (RHS) of  \eqref{eq:condTSCaffpspec} increases in $\rho$; moreover, the probability of success in Thm.~\ref{thm:TSCprob} increases in $n$. If the number of subspaces, $L$, increases, for fixed $d,\q$ and $n$, clustering intuitively becomes harder and, indeed, the RHS of \eqref{eq:condTSCaffpspec} is seen to decrease in $L$. Note that Thm.~\ref{thm:TSCprob} does not apply to subspaces that intersect as $\affp(\cS_k,\cS_l)\!=\!1$ if $\cS_k$ and $\cS_l$ intersect and the RHS of \eqref{eq:condTSCaffpspec} is strictly smaller than $1$. 
We next present a result analogous to Thm.~\ref{thm:TSCprob} that applies to intersecting subspaces. 


\vspace{-0.15cm}
\begin{theorem}
Suppose $n=\rho \q$ data points are chosen in each of the $L$ subspaces at random according to $\vx_j^{(l)} = \mU^{(l)} \va^{(l)}_j, j =1,..., n$, where the $\va^{(l)}_j$ are i.i.d.~uniform on $\US{d}$ and $\rho \geq 6$. If 
\vspace{-0.3cm}
\begin{align}
\max_{k\neq l} \aff(\cS_k,\cS_l) \leq   \frac{1}{13 \log N } ,
\label{eq:condTSCaff2}
\end{align}
 then the TSC subspace detection property holds with probability at least
$
1 - 3/N - Ne^{-c n}
$,
\label{thm:aff2}
where $c$ is an absolute constant.
\end{theorem}
\vspace{-0.2cm}
The interpretation of Thm.~\ref{thm:aff2} is analogous to that of Thm.~\ref{thm:TSCprob} with the important difference that the RHS of \eqref{eq:condTSCaff2}, as opposed to the RHS of \eqref{eq:condTSCaffpspec},  decreases, albeit slowly, in $n$ (recall that $N=Ln$). For SSC a result in the flavor of Thm.~\ref{thm:aff2} was reported in \cite[Thm.~2.8]{soltanolkotabi_geometric_2011}.

\vspace{-0.55cm}
\section{Erasures}
\vspace{-0.35cm}

In practical applications the data points to be clustered are often corrupted by erasures, e.g., images that need to be clustered could exhibit scratches. 
Understanding the impact of erasures on clustering performance is obviously of significant importance. The literature seems, however,  essentially void of corresponding analytic results. 
In the deterministic subspace setting such results will necessarily depend on the specific orientations of the subspaces. 
In the following, we therefore take both the orientations of the subspaces and the points in each subspace to be random. Specifically, we take the entries of the $\mU^{(l)} \in \reals^{m\times d}$ to be i.i.d.~$\mathcal N(0,1/m)$, which ensures that each of the $\mU^{(l)}$ is approximately orthonormal with high probability. 
\vspace{-0.1cm}
\begin{theorem}
Suppose $n = \rho \q$ data points are chosen in each of the $L$ subspaces at random according to $\vx_j^{(l)} = \mU^{(l)} \va^{(l)}_j, j =1,..., n$, where the $\va^{(l)}_j$ are i.i.d.~$\mathcal N(0, (1/d) \, \mI)$ and $\rho \geq 10/3$. 
Assume that in each $\vx_j$ up to $s$ entries (possibly different for each $\vx_j$) are erased, i.e., set to $0$. 
Let the entries of each matrix $\mU^{(l)} \in \reals^{m\times d}$ be i.i.d.~$\mathcal N(0,1/m)$. If 
\[
m\geq  c_2  \frac{\sqrt{\log N}}{\sqrt{\log \rho}} 
 \left(   d \log\left( c_4 \frac{\sqrt{\log N}}{\sqrt{\log \rho}}  \right) \!+\! s \log\left( \frac{me}{2s} \right)  \!+ \! \log L     \right) \!+\! c_0 s,
\]
then the TSC subspace detection property
holds with probability at least
$
1 - L n   e^{-c_3 n}  - \frac{L}{(L-1)^3 n^2 }     - 4e^{-c_1 m},
$
where $c_0,c_1,c_2,c_3,c_4>0$ are absolute constants. 
\label{thm:fullyrandomnew}
\end{theorem}
\vspace{-0.1cm}
Strikingly, Thm.~\ref{thm:fullyrandomnew} shows that the number of erasures is allowed to scale (up to a log-factor) linearly in the ambient dimension. 

For the fully random data model used in Thm.~\ref{thm:fullyrandomnew} we can furthermore conclude that 
TSC succeeds with high probability even when the dimensions of the subspaces scale (up to a log-factor) linearly in the ambient dimension. 
Drawing such a conclusion from Thm.~\ref{thm:TSCprob} or Thm.~\ref{thm:aff2} seems difficult as the relation between $m,d$, and $L$ is implicit in the affinity measures. These findings should, however, be taken with a grain of salt as the fully random subspace model ensures that the subspaces are approximately disjoint with high probability. 
In the erasure-free case, i.e., for $s=0$, a result for SSC,  analogous to Thm.~\ref{thm:fullyrandomnew}, was reported in \cite[Thm.~1.2]{soltanolkotabi_geometric_2011}.

\vspace{-0.2cm}
\section{Detection of outliers}
\vspace{-0.3cm}
\label{sec:detoutl}
Outliers are data points that do not lie in one of the low-dimensional subspaces $\cS_l$ and have no low-dimensional linear structure. Here, this is modeled by assuming random outliers distributed uniformly on the unit sphere of $\reals^m$. The outlier detection criterion we employ does not need knowledge of the number of outliers $N_0$ and is based on the following observation. The maximum inner product between an outlier and any other point in $\X$ is, with high probability, smaller than 
$
c  \sqrt{\log N}/\sqrt{m}.
$
We therefore classify $\vx_j$ as an outlier if 
$
\max_{p\neq j} \left|\innerprod{\vx_p}{\vx_j} \right| < c \sqrt{\log N} / \sqrt{m}. 
$
The maximum inner product between any point $\vx_j \in \X_l$ and the points in $\X_l \setminus \vx_j$ is unlikely to be smaller than $1/\sqrt{d}$. 
Hence an inlier is unlikely to be misclassified as an outlier if $d/m$ is sufficiently small. 
\vspace{-0.1cm}
\begin{theorem}
Suppose $n = \rho \q$ data points are chosen in each of the $L$ subspaces at random according to $\vx_j^{(l)} = \mU^{(l)} \va^{(l)}_j, j =1,..., n$, where the $\va^{(l)}_j$ are i.i.d.~uniform on $\US{d}$ and each $\mU^{(l)}$ is orthonormal. Let the $N_0$ outliers be i.i.d.~uniform on $\US{m}$. 
Declare $\vx_j \in \X$ to be an outlier if
$
\max_{p\neq j} \left| \innerprod{\vx_p}{\vx_j}\right| <  \sqrt{ 6 \log N} / \sqrt{m}
$. 
Then, with $N=Ln + N_0$, provided that
\vspace{-0.2cm}
\begin{align}
\frac{d}{m} \leq \frac{1}{6 \log N}
\label{eq:condoutldet}
\end{align}
\vspace{-0.4cm}

\noindent with probability at least
$
1-2N_0/N^2 - nL e^{-\log(\pi/2) (n-1)},
$
every outlier is detected and no point in a subspace is misclassified as an outlier.
\label{thm:outldete}
\end{theorem}
\vspace{-0.2cm}
Since \eqref{eq:condoutldet} can be rewritten as
$
N_0 \leq e^{\frac{m}{6d}} - L n,
$
we can conclude that outlier detection succeeds, even if the number of outliers scales exponentially in $m/d$, i.e., if $d$ is kept constant, exponentially in the ambient dimension! Note that this result does not make any assumptions on the orientations of the subspaces $\cS_l$. 
The outlier detection scheme proposed in \cite{soltanolkotabi_geometric_2011} allows to identify outliers under a very similar condition. However, it requires the solution of $N$ $\ell_1$-minimization problems, each in $N$ unknowns, while the algorithm proposed here needs to compute $N^2$ inner products followed by thresholding only.

\vspace{-0.4cm}
\section{Numerical results}
\vspace{-0.3cm}
We use the performance measures employed in \cite{soltanolkotabi_geometric_2011,elhamifar_clustering_2010}. 
 The \emph{clustering error} (CE) is defined as the ratio between the number of misclassified data points and the total number of points in $\X$.
 The error in estimating the number of subspaces $L$ is denoted as EL and takes on the value $0$ if the estimate is correct, else it is equal to $1$. 
 The \emph{feature detection error} (FDE) is defined as
$
\frac{1}{N} \sum_{i=1}^N \left(1 - \norm[2]{\vb_{\vx_i}} /\norm[2]{\vb_i}   \right),
\label{eq:fde}
$
where $\vb_i$ is the $i$th column of the adjacency matrix $\mA$ and $\vb_{\vx_i}$ is the vector containing the entries of $\vb_{i}$ corresponding to the subspace $\vx_i$ lives in. 
The FDE measures to which extent points from different subspaces are connected in $G$ 
and is equal to zero if the TSC subspace detection property holds. 

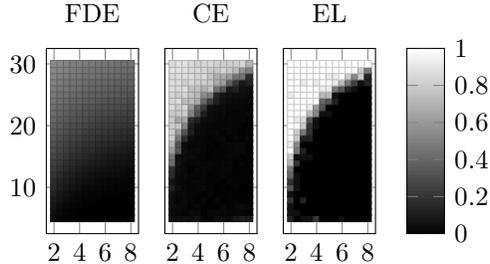
\begin{figure}
\centering
\begin{tikzpicture}[scale=1]
\begin{groupplot}[group style={group size=3 by 1,horizontal sep=0.35cm,vertical sep=0.2cm,xlabels at=edge bottom, ylabels at=edge left,yticklabels at=edge left},
width=0.17\textwidth,height=0.245\textwidth,
/tikz/font=\small, colormap/blackwhite, point meta min = 0, point meta max=1]
    \nextgroupplot[title={FDE }]
     \addplot[mark=square*,only marks, scatter, scatter src=explicit,
     mark size=1.3, colormap/blackwhite,
     domain=0:1]
      file {./FDE_varyd.dat};

  \nextgroupplot[title={CE}]
     \addplot[mark=square*,only marks, scatter, scatter src=explicit,
     mark size=1.3, colormap/blackwhite]
 file {./CE_varyd.dat};

\nextgroupplot[title={EL}, colorbar]
      \addplot[mark=square*,only marks, scatter, scatter src=explicit,
  mark size=1.3,colormap/blackwhite]
   file {./LE_varyd.dat};
  \end{groupplot}
\end{tikzpicture}

\vspace{-0.4cm}
\caption{\label{fig:varyd}Errors as a function of the dimension of the subspaces, $d$, on the vertical and $\rho$ on the horizontal axis.}
\vspace{-0.3cm}
\end{figure}

\begin{figure}
\centering
\begin{tikzpicture}[scale=1]
\begin{groupplot}[group style={group size=4 by 1,horizontal sep=0.15cm,vertical sep=0.0cm,xlabels at=edge bottom, ylabels at=edge left,xticklabels at=edge bottom,yticklabels at=edge left},
width=0.18\textwidth,height=0.177\textwidth,/tikz/font=\small, colormap/blackwhite],point meta min = 0, point meta max=0.8]

    \nextgroupplot[title={$s=0$}]
     \addplot[mark=square*,only marks, scatter, scatter src=explicit,
     mark size=1.5,colormap/blackwhite]
     file {./CE_EF0.dat};

  \nextgroupplot[title={$s=5$}]
     \addplot[mark=square*,only marks, scatter, scatter src=explicit,
     mark size=1.5,colormap/blackwhite]
     file {./CE_EF5.dat};

  \nextgroupplot[title={$s=10$}]
     \addplot[mark=square*,only marks, scatter, scatter src=explicit,
     mark size=1.5,colormap/blackwhite]
     file {./CE_EF10.dat};

  \nextgroupplot[title={$s=15$},colorbar]
     \addplot[mark=square*,only marks, scatter, scatter src=explicit,
     mark size=1.5,colormap/blackwhite]
     file {./CE_EF15.dat};

  \end{groupplot}
\end{tikzpicture}
\caption{\label{fig:varydsc}CE as a function of the dimension of the subspaces, $d$, on the vertical and $\rho$ on the horizontal axis.}
\vspace{-0.5cm}
\end{figure}
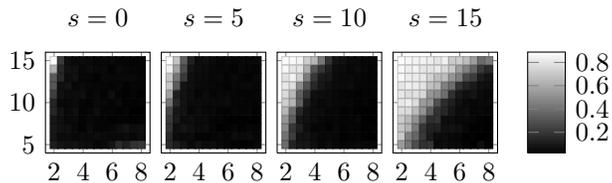

{\bf Influence of $d$, $\rho$, and erasures:} We generate $L=15$ subspaces in $\reals^{50}$ at random, by choosing  the corresponding $\mU^{(l)}$ uniformly at random from the set of orthonormal matrices in $\reals^{m\times d}$, and vary the number of points $n=d \rho$ in each subspace. The points in the subspaces are chosen at random according to the probabilistic model in Thm.~\ref{eq:condTSCaff2}. 
The results 
depicted in Fig.~\ref{fig:varyd} show, as indicated in Sec.~\ref{sec:intro}, that TSC can, indeed, succeed even when the TSC subspace detection property does not hold. 
Finally, we perform the same experiment, but erase the entries of $\vx_i$ with indices in $\D_i$, where $\D_i$ is chosen independently for each $\vx_i$ and uniformly from $\{\D\subseteq \{1,...,m\}\colon |\D|=s\}$. The results summarized in Fig.~\ref{fig:varydsc} show that TSC succeeds, even when a large fraction of the entries is erased. 

{\bf Detection of outliers:}
In order to allow for a comparison with the outlier detection scheme proposed in \cite{soltanolkotabi_geometric_2011}, we perform our experiment with the same parameters as used in \cite[Sec. 5.2]{soltanolkotabi_geometric_2011}. 
Specifically, we set $d=5$, vary $m=\{50,100,200\}$, and generate $L=2 m/d$ subspaces and $n=5d$ points in each subspace at random as in the previous paragraph. 
Each of the $N_0 = Ln$ outliers is chosen i.i.d.~uniformly on $\US{m}$. Note that we have as many outliers as inliers. We find a
  misclassification error probability of $\{0.017, 1.5 10^{-4}, 2.5 10^{-5}\}$ for $m=\{50,100,200\}$, respectively. 
Similar performance was reported for the scheme proposed in \cite{soltanolkotabi_geometric_2011}.

%

%






\end{document}